\documentclass[aps,pre,twocolumn,10pt,amsmath,amsfonts,amssymb,floatfix,bibnotes]{revtex4-1}
\usepackage{amssymb}
\usepackage{graphicx,color}
\usepackage{bbm}
\usepackage{multirow}

\usepackage{amsmath,bm,epsfig}
\newcommand{\B}[1]{{\bm{#1}}}

\hypersetup{
pdfauthor={Giorgio Parisi, Yoav G. Pollack, Itamar Procaccia, Corrado Rainone and Murari Singh},
pdftitle={How Generic are the Robust Theoretical Aspects of Jamming in Hard Sphere Models?},
pdfsubject={Soft Condensed Matter},
pdfkeywords={Amorphous, Soft Matter, force\-law, hard spheres, jamming, colloids, simulation, Emergent interactions},}

\begin{document}

\title{How Generic are the Robust Theoretical Aspects of Jamming in Hard Sphere Models?}

\author{Giorgio Parisi$^1$}
\author{Yoav G. Pollack$^2$}
\author{ Itamar Procaccia$^2$}
\author{ Corrado Rainone$^2$}
\author{ Murari Singh$^2$}
\affiliation{$^1$Dipartimento di Fisica, Sapienza Universit`a di Roma, INFN, Sezione di Roma I,	IPFC – CNR, Piazzale Aldo Moro 2, I-00185 Roma, Italy.\\
$^2$Dept of Chemical Physics, The Weizmann Institute of Science, Rehovot 76100, Israel.}

\begin{abstract}
In very recent work the mean field theory of the jamming transition in infinite dimensional hard spheres
models was presented. Surprisingly,  this theory predicts {\em quantitatively} numerically determined characteristics of jamming
in two and three dimensions. This is a rare and unusual finding. Here we argue that this agreement in non-generic: only for hard sphere models it happens
that sufficiently close to jamming the effective interactions are in agreement with mean-field theory, justifying the truncation of
many body interactions (which is the exact protocol in infinite dimensions). Any softening of the {\em bare} hard sphere interactions results
in {\em effective} interactions that are not mean-field all the way to jamming, making the discussed phenomenon non generic.
\end{abstract}
\maketitle

Models of hard sphere fluids and solids provided useful insights in condensed matter physics and in statistical mechanics for many decades \cite{06HM,99Rap}. In the last two decades hard spheres played a particularly important role in the investigation of the jamming transition, modeling the solidification of compressed granular matter \cite{03OSLN}. The jamming transition is a critical phenomenon, characterized by a number of critical exponents whose values are typically irrational. Careful numerical simulations in 2 and 3 dimensions could determine
these exponents quite precisely, with three to five digits accuracy \cite{02SEGHL,02OLLN,16BCNO}. Direct theoretical calculations in finite dimensions are
not available, but a theory in infinite dimensions is available, providing exact predictions as $d\to \infty$ for these
critical exponents. It turned out that the predictions at $d\to \infty$ appear to actually agree quantitatively with simulations
results in $d=2$ and 3 \cite{14CKPUZ,15CCPZ,11BJZ}. In the words of Ref.~\cite{17CKPUZ},
``{\em One  of  the  most  remarkable  features  of  the $d\to \infty$ solution  is  its  agreement  with  both
qualitative and quantitative aspects of jamming observed in numerical simulations.  This
outcome is especially stunning...."}. Indeed stunning, and highly unusual: the critical exponents are usually strongly dependent
on dimension, and in many cases turn into mean-field values above some critical dimension. For the jamming problem it was found that
non-trivial exponents are $d$-independent from $d=2$ to $d\to \infty$. The aim of this Letter is to explain this unusual phenomenon and to argue that it is non-generic, being fragile to any degree of softening of the hard sphere potential. The proposed answer is simple: the theory in infinite dimension is a perturbative approach that enjoys important simplifications by neglecting higher order terms that are hard to deal with in finite dimensions \cite{06PZ}. Below we demonstrate that near jamming, this is also the situation for hard spheres in 2 and 3 dimensions \cite{06BW}, but only for hard spheres. Softening the hard sphere potential
introduces unavoidable complications that mar the correspondence between low and infinite dimensions.

A key concept that underlies the discussion below is that of {\em effective forces}. These must be distinguished from the {\em bare forces}. For example in hard spheres the bare forces are zero when there is no contact and infinity upon contact.
In a thermal ensemble particles collide and impart momenta, from which one can compute the effective forces \cite{06BW}. If the bare potential is a function of the distance between particles, the effective force can be measured simply as an integral over the force,
divided by the total interaction time. Another way to determine the effective forces, which
is only appropriate for a glassy system, is to compute the time-averaged positions of the particles $\{\bar {\B r_i}\}_{i=1}^N$, and ask which effective
forces stabilize these averaged positions to make them time independent. The second method  \cite{16GLPPRR,16GPP} is more appropriate for
experiments in which the momentum transfer or the actual time dependent bare force are hard to measure. In the simulations below
we employ the first method.

To find the effective interactions in hard disks we executed event driven 2-dimensional simulations for systems with fixed area $A$, a given number of disks $N=400$ at temperature $T=1$. The area is square with periodic boundary conditions and all the disks have the same mass $m$. The disk radii $R$ are slightly poly-dispersed around a binary 50:50 distribution with mean values and standard deviations of
$\langle R_A\rangle = 0.5$, $\sigma_{R_A} = 0.0081$, $\langle R_B\rangle = 0.7$ and $\sigma_{R_B} = 0.0123$. All the hard disk simulations
started by determining accurately the area $A_{\rm in} \equiv L_{\rm in}^2$ for which the system jams,
and see Supplemental Material. Then we choose a density $\rho$ for the unjammed systems by expanding
the system size according to
\begin{equation}
L=L_{\rm in} \times (1+\epsilon) \ .
\label{defroho}
\end{equation}
After expanding the system size from the jammed state by the desired $\epsilon$, the simulation is first equilibrated for $2\times 10^6$ collisions and then run for $10^8$ collisions. This long run is examined manually to ascertain that it has a section of at least $n_c=10^7$ collisions in which there are no transitions between meta-basins. Disk positions were measured and averaged-over with the maximal resolution possible (i.e. after each collision). Given a value of $\epsilon$ the effective forces $\B f_{ij}$ were measured by computing the momentum
transfer $\Delta_k \B p(i,j)$  during every $k'th$ collision between particles $i$ and $j$, followed by averaging according to
\begin{equation}
\B f_{ij} \equiv \frac{\sum_k \Delta_k \B p(i,j)}{t} \ , \quad \text{for hard spheres}
\label{fijmom}
\end{equation}
where $t $ is the total duration of the measurement.
One test of the accuracy of the forces $\B f_{ij}$ is the requirement that they uphold force balance, or
\begin{equation}
\B f_i=\sum_j \B f_{ij} = 0 \ .
\label{sumrule}
\end{equation}
For hard disks this sum rule was obeyed in our numerics to better than 1 part in $10^3$ in units of the mean inter-particle force. The reader should notice that although we measure $\B f_{ij}$ by tracking {\bf interactions} between particles $i$ and $j$ it is {\bf  not} generally a function of only $\bar {\B r}_{ij} \equiv \bar {\B r}_i-\bar{\B  r}_j$. Even in mean-field theory one expects that the effective forces between particles $i$ and $j$ will depend
on some characteristic of the cages in which they move, in addition to $\bar {\B r}_{ij}$. In correspondence
with the mean field theory of spin glasses we can expect a dependence on the mean-square fluctuations
in the respective two cages. Defining $K_i\equiv \sqrt{\langle ( \B r_i(t)-\bar {\B r}_i)^2\rangle}$ we expect that \cite{82P,91GY}\footnote{In principle one can also consider a dependence on a tensorial quantity $K^{\alpha\beta}_i\equiv \sqrt{\langle ( r^\alpha_i(t)-\bar {r}^\alpha_i)( r^\beta_i(t)-\bar {r}^\beta_i)\rangle}$}
\begin{equation}
\B f_{ij} =g(K_i, K_j,  {\B {\bar r_{ij}}})   \quad \text{in mean field theory} \ ,
\label{MF}
\end{equation}
with $g$ being an a-priori unknown function.
In finite dimensions multiple collisions and the effect of successive collisions cause the effective forces to depend also on the averaged positions of other particles. In other words the effective forces are not in general mean-field, and see below for more details. We will show however that for hard disks sufficiently close to jamming the effective forces are binary, and even
independent of $K_i$ and $K_j$.

\begin{figure}
\includegraphics[scale = 0.38]{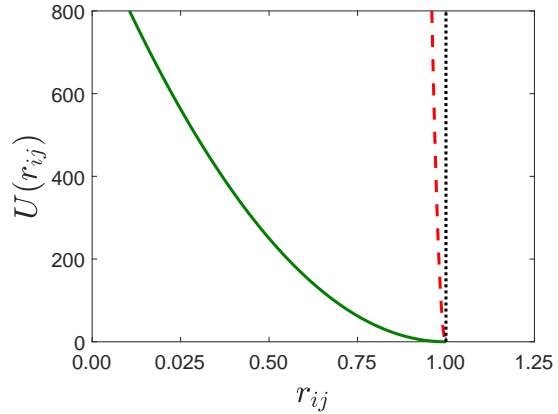}
	\caption{Comparison between the hard and the soft potentials used in this Letter to underline the fragility of
the hard sphere limit. The vertical black line is the hard disk potential and the two others illustrates Eq.~(\ref{softpot}) with
$V_0=500000$ (red dashed line) and $V_0=1000$ (green continuous line).}
\label{compare}
\end{figure}

To underline the difference between hard spheres and more generic potentials we study here a system of softer disks \cite{03OSLN}. Here the system consists
of a 50:50 mixture of `small' (A) and `large' (B) particles, with diameter ratio of $\lambda_B/\lambda_A=1.4$. This system has been used extensively in numerical simulations to create jammed configurations. The {\em bare} interaction potential between particles $i$ and $j$
is given by \cite{11SOS}
\begin{equation}
U(r_{ij})=\frac{V_0}{2}\left( 1-\frac{r_{ij}}{\lambda_{ij}}\right)^2\Theta\left( 1-\frac{r_{ij}}{\lambda_{ij}}\right)\ ,
\label{softpot}
\end{equation}
where $r_{ij}$ is the distance between the center of masses of the particle $i$ and $j$, $\lambda_{ij}=(\lambda_i+\lambda_j)/2$ is the average diameter,
$V_0$ is strength of interaction, and $\Theta$ is the Heaviside step function. To remain close to the hard sphere limit we choose first a large value of $V_0=500000$. A second comparison to a softer interaction is achieved with $V_0=1000$. The comparison between the hard sphere limit and these two softer potentials is shown in Fig.~\ref{compare}.
The units for energy and length are $V_0$ and $\lambda_A$ respectively. We integrate the equations of motion for this system
using a standard  velocity-Verlet  algorithm with time step $\Delta t=0.001$ for $V_0=1000$ and  $\Delta t=0.0002$ for $V_0=500000$. The Nose-Hoover chain thermostat was used to maintain the desired temperature. After expanding the system size from the jammed state by the desired $\epsilon$, we first equilibrate the system for time $\tau_1=10^5$ (in reduced units) (For more details see Supplemental Material). The simulation is then run at constant $T=10^{-3}$ for a further time of $\tau_2=10^6$. All the results pertaining to the soft disks are extracted as described above from a time segment of length $\tau_2/10$. This is again done to avoid  transitions between meta-basins. The upshot of this choice is that the average positions were determined using averaging times that are well below the time for which particle diffusion destroys their meaning. In practice this constrains the values of the expansion $\epsilon$. For values of $\epsilon\geq 5.5\times 10^{-2}$ for hard disks and $\epsilon > 10^{-2}$ for soft disks we could not determine the mean positions of the particles with sufficient accuracy. The average position of the particles are denoted as above as $\{\bar {\B r}_i\}_{i=1}^N$. In the present case the effective forces are computed from the dynamics according to
\begin{equation}
\B f_{ij} \equiv - \frac{1}{t}\int_0^t dt' \left(\frac{\partial U}{\partial \B r_{ij}}\right) \ ,
\quad
\begin{array}{c}
\text{for differentiable}\\
\text{ bare potential}\
\end{array}
,
\label{def2}
\end{equation}
with $t$ being the time of integration of the dynamics. The sum rule (\ref{sumrule}) was well obeyed for the soft disks to better than 1 part in $10^6$ (in units of the mean inter-particle force). Again the main question below will be whether these effective forces are mean field
as in Eq.~(\ref{MF}), or whether they exhibit many-body interactions.
\begin{figure}
\includegraphics[scale = 0.36]{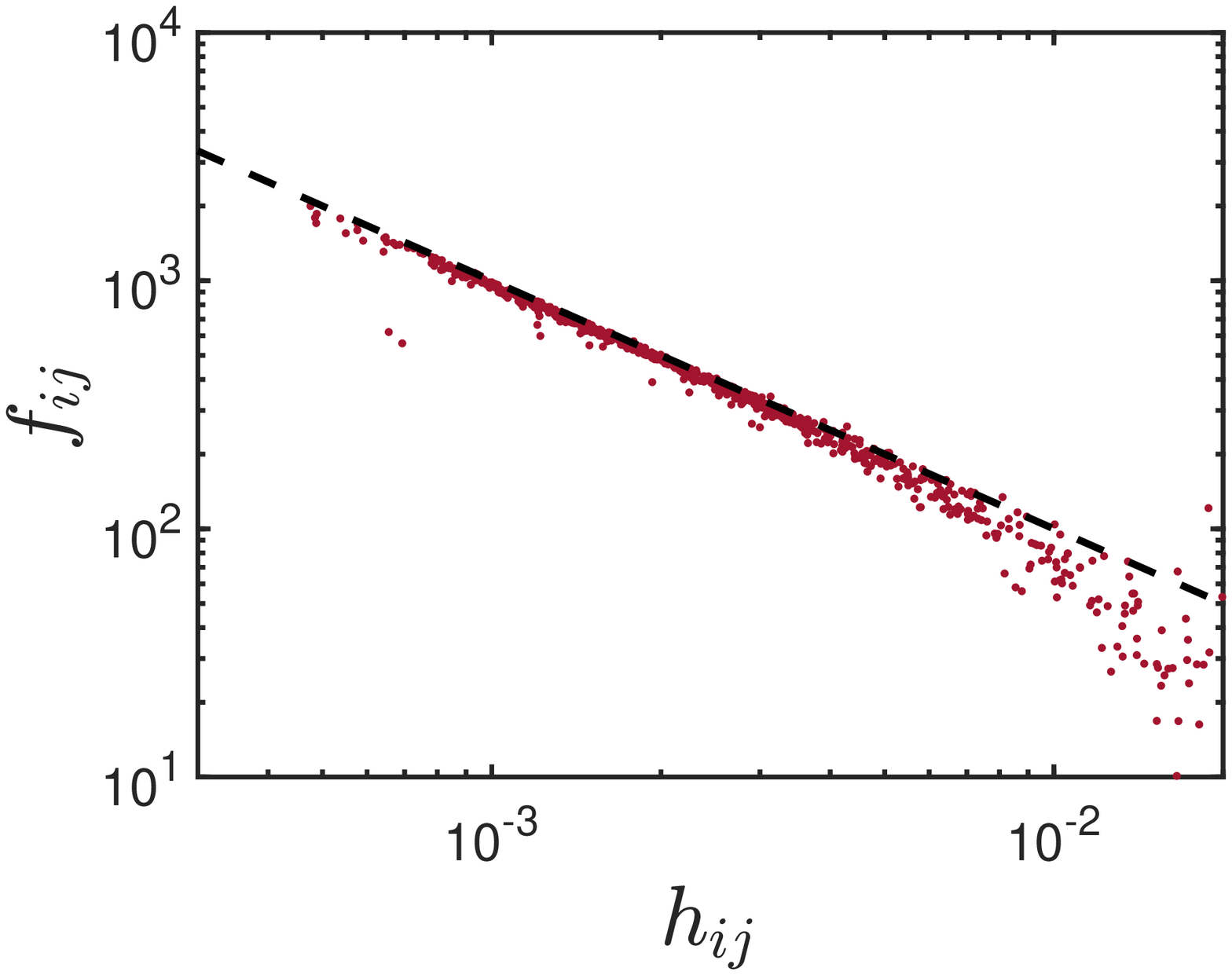}
\includegraphics[scale = 0.50]{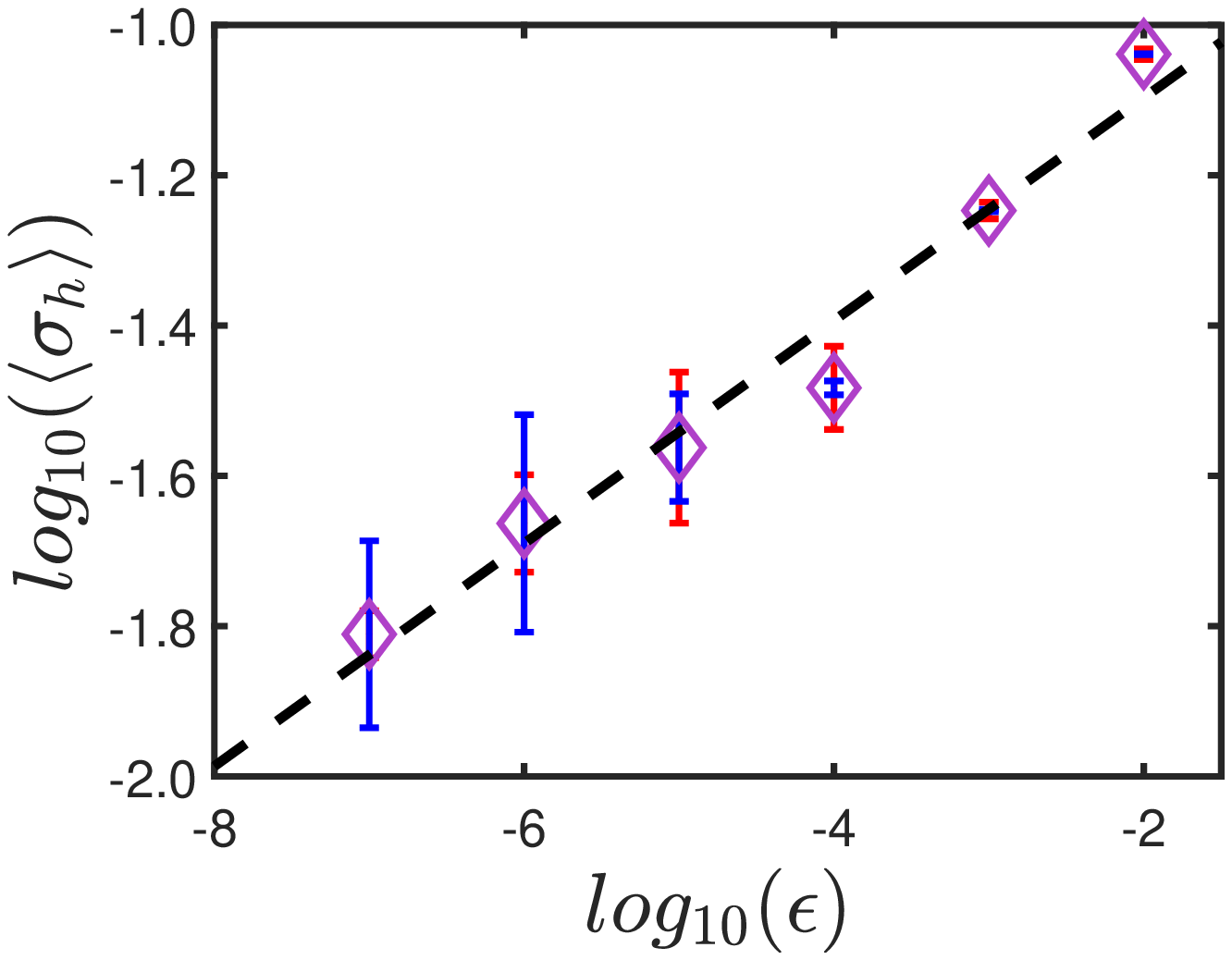}
\caption{Upper panel: raw data of $f_{ij}$ vs. $h_{ij}$ at $\epsilon=10^{-3}$. The 2-body force-law (\ref{thij}) is represented
by the dashed line. Note that the scatter is not due to inaccuracy as can be tested by the high precision with which Eq.~(\ref{sumrule}) is obeyed. Lower panel: The standard deviation from the two-body force law for hard disks as a function of the distance from jamming. In dashed line we present the power-law fit (\ref{powerlaw}).}
\label{fijhard}
\end{figure}

The analysis is easier in the case of hard disks with which we begin.
In this case we find that near jamming the effective forces trivialize to binary interactions. To show this we follow two steps: (i) finding for each configuration a function of $\B {\bar r_{ij}}$  that fits best the effective forces $\B f_{ij}$; then (ii)  measuring the deviations of the data from this best function. The point is that when the effective forces $\B f_{ij}$ are purely 2-body forces they should be a function of $\B {\bar r_{ij}}$ with a scatter that is only allowed
 by the accuracy of the measurement, which is extremely high as can be tested by the agreement with Eq.~(\ref{sumrule}).  On the other hand when the effective forces include many-body corrections the data should scatter around the best function, with the degree of scatter proportional to the relative significance of the many-body forces.

For hard disks the only energy is $T$ and the only typical scales are $h_{ij}$,  the average gaps between particles \footnote{see in the Supplemental Material another, scalar definition of the average gap and its consequences}
\begin{equation}
h_{ij} \equiv |\bar r_{ij} - R_i-R_j| \ ,
\label{defh}
\end{equation}
one can expect that if the effective forces were exactly binary than \cite{06BW}
\begin{equation}
f_{ij}=T/h_{ij} \ , \quad \text{for hard disks} \ .
\label{thij}
\end{equation}
In Fig.~\ref{fijhard} upper panel we show typical data for $f_{ij}$ as a function of $h_{ij}$ for $\epsilon=10^{-3}$. We see that the data do not form a function.
Next we calculated standard deviation $\sigma_h(\epsilon)$ around the compensated data $f_{ij}\times h_{ij}/T$. This quantity was averaged over different initial configurations for each expansion $\epsilon$. The lower panel of Fig.~\ref{fijhard}  shows  $\sigma_h$ for different values of $\epsilon$ in a log-log plot. The red error bars in this figure stem from standard deviation between different configurations. The blue error bars represent the accuracy in determining the inter-particle forces from Eq.~(\ref{sumrule}).
The deviation from purely binary interactions decreases upon approaching jamming:
\begin{equation}
\langle \sigma_h(\epsilon)\rangle \sim \epsilon^\zeta \ , \quad \zeta\approx 0.15\pm 0.04  \ .
\label{powerlaw}
\end{equation}
It remains to be seen whether this critical exponent can be derived from known exponents of the jamming criticality or is
this a new exponent for the problem at hand.
\begin{figure}
\includegraphics[scale = 0.50]{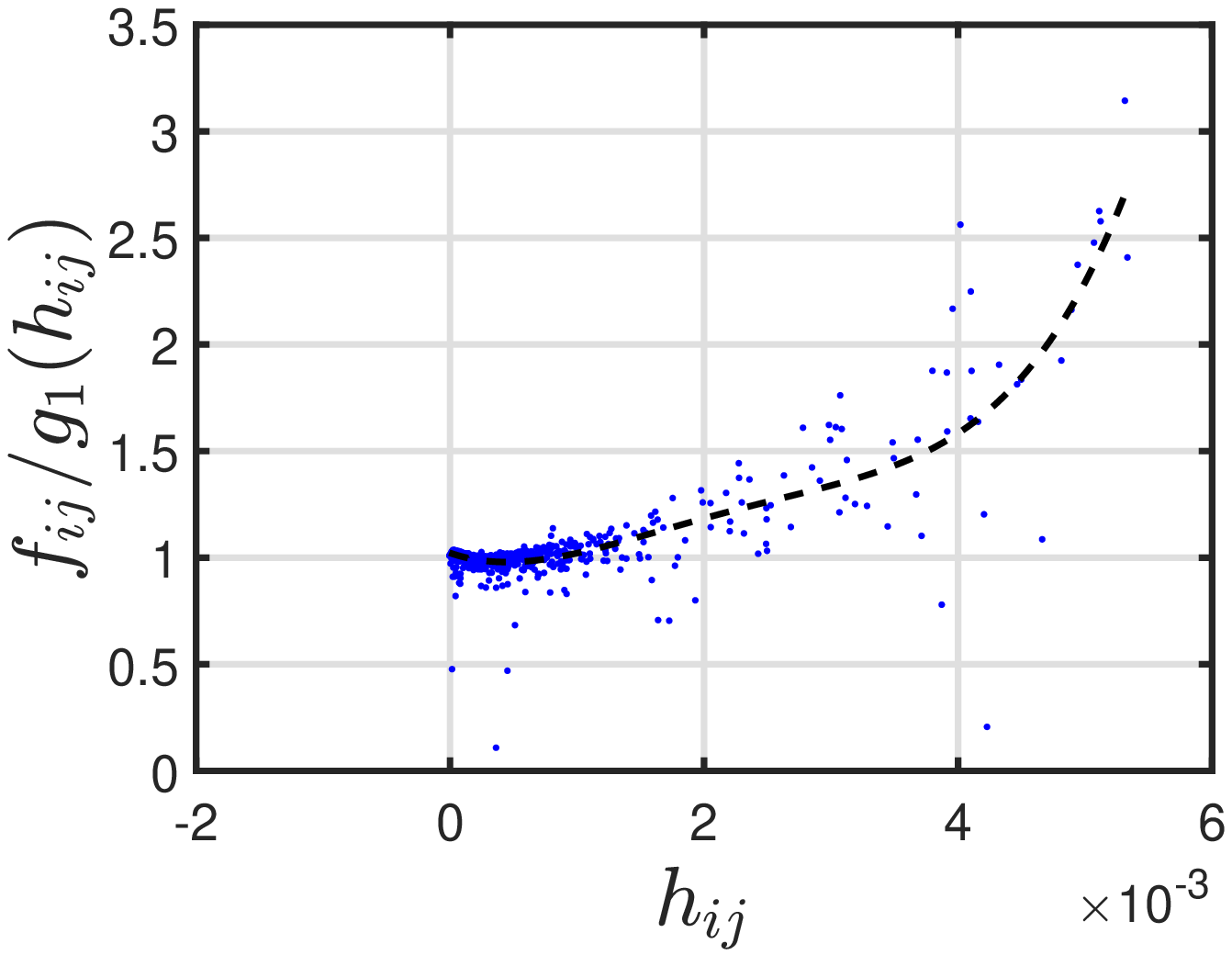}
	\includegraphics[scale = 0.50]{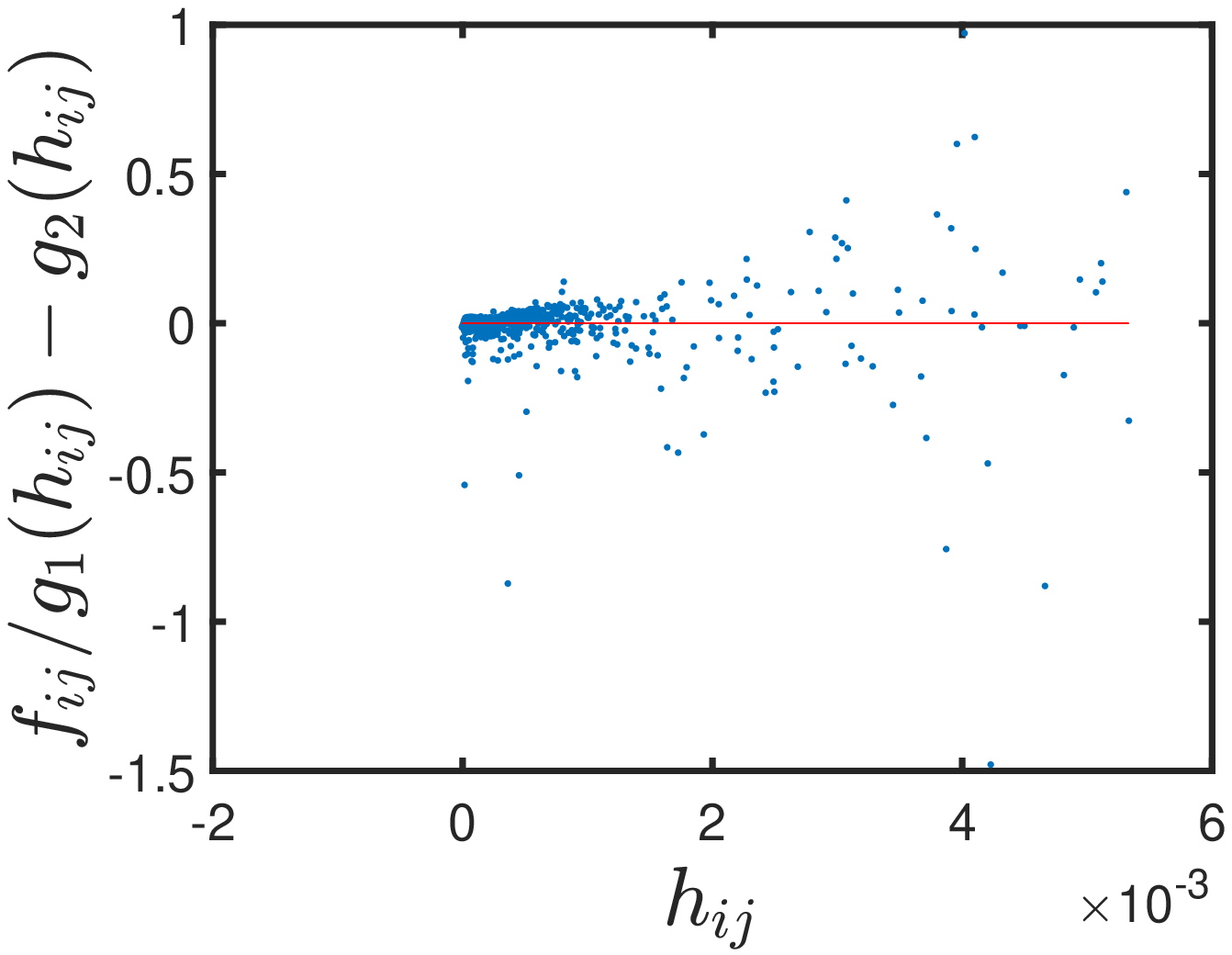}
	\caption{The procedure followed to determine the deviations from mean field binary force laws
in the case of soft spheres with $V_0=500000$  and $\epsilon=10^{-4}$. Upper panel: the first step after normalizing by the Pad\'e approximant. Black dashed line indicates the $g_2(h_{ij})$ fit.
Lower panel: the second step resulting in a scatter around zero. Note that for softer spheres with $V_0=1000$ the
first step with a Pad\'e approximation is sufficient in many instances.}
\label{procedure}
\end{figure}
\begin{figure}
\includegraphics[scale = 0.50]{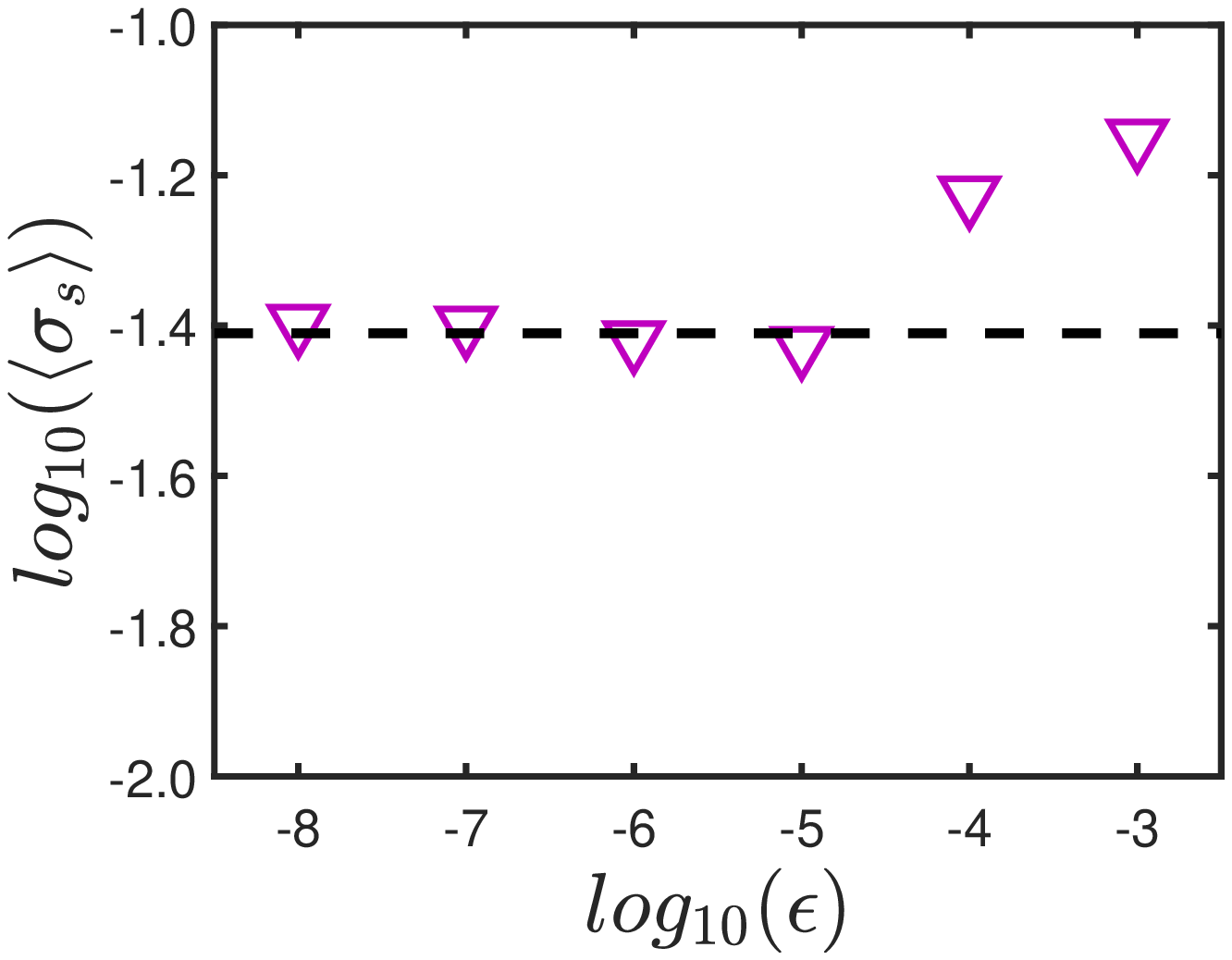}
	\includegraphics[scale = 0.50]{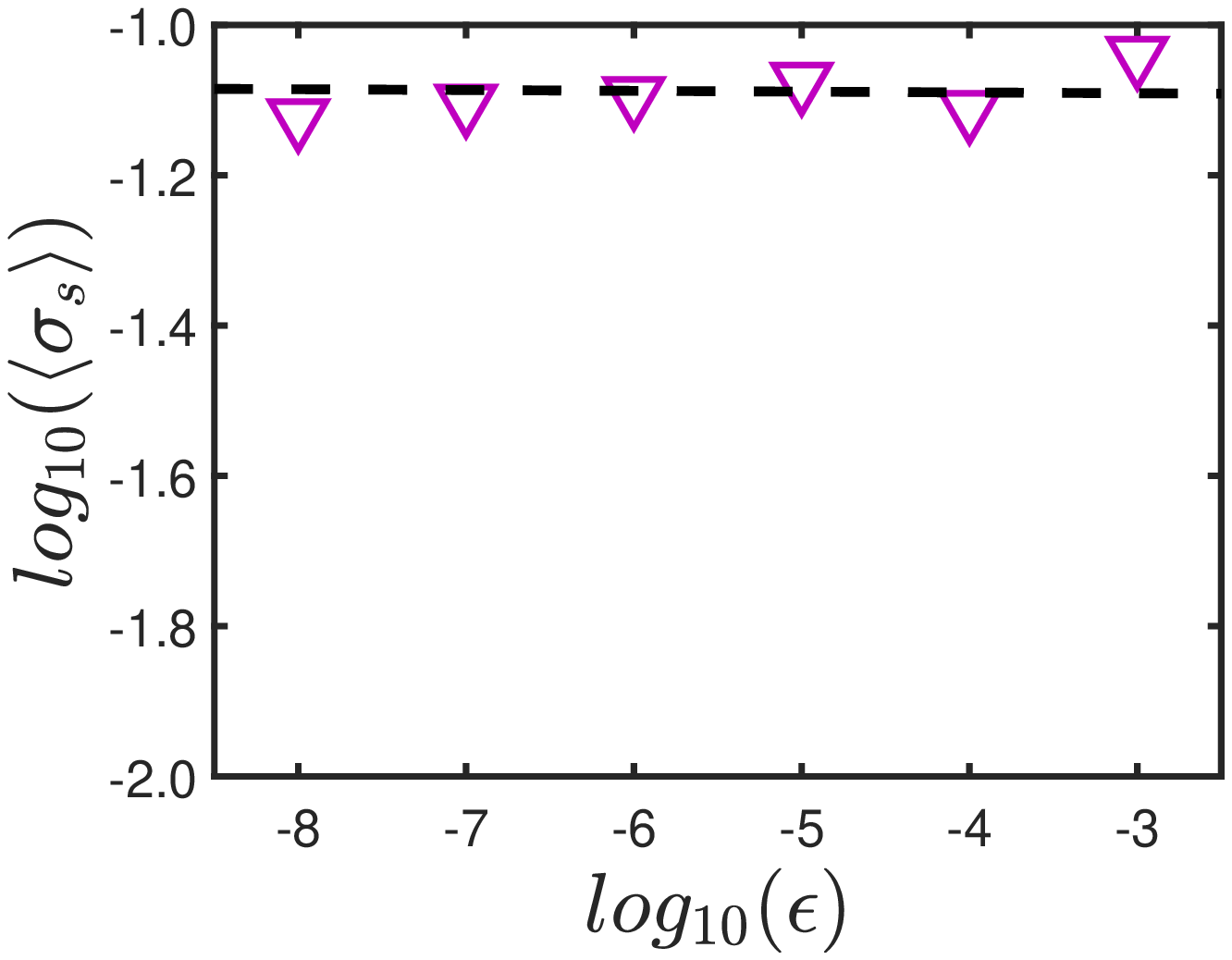}
	\caption{The standard deviation from the mean field binary force law for soft spheres as a function of the distance from jamming.
Upper panel: $V_0=500000$. Lower panel $V_0=1000$.}
	\label{fijsoft}
\end{figure}

In the case of soft disks, we have more length scales $\lambda_{ij}$ to deal with, and guessing the form of a putative limiting effective interaction becomes less obvious.
In addition we need now to consider the 3 different interactions AA,AB,BB. Naive fitting choices like polynomial and rational fits proved not to be accurate enough for our purpose (as detailed below). This called for a somewhat more complicated procedure as explained next.

In order to determine whether in a given system the force-law conforms with mean-field expectation Eq.~\ref{MF} we need first to determine the cage fluctuations $K_i$.
The probability distribution function (pdf) of $K_i$ was measured for soft spheres using some 77-92 configurations (depending on $\epsilon$). Selecting only pairs of particles with very close values $K_i\approx K_j$ from the peak of this distribution,
we used a Pad\'e approximant (of order 1 in the numerator and order 2 in the denominator) to fit the forces $f_{ij}$ as functions of
$h_{ij}\equiv {\bar r_{ij}}-\lambda_{ij}$ . The function obtained is denoted as $g_1(h_{ij})$.
Second, we normalized $f_{ij}$ by this function. Doing this, one discovers that  the Pad\'e approximant is not always sufficient, in the sense  that $f_{ij}/g_1$  still retains a clear functional form deviating from unity, see top panel of Fig.~\ref{procedure}.  We therefore fitted to the
 normalized forces a polynomial of degree 4. The new fit is denoted as $g_2(h_{ij})$.  Next we considered, as seen in Fig.~\ref{procedure}, the function $f_{ij}/g_1-g_2$ and determined the standard deviation $\sigma_s$ of the data scatter around zero. Finally we averaged $\sigma_s$ among the 3 interactions, for each expansion parameter $\epsilon$ \footnote{Note that in the case of soft spheres
we computed the standard deviation from all the samples together, whereas in the hard spheres
we averaged over individual configurations. Doing the same in the present introduces no difference.}. Fig.~\ref{fijsoft} shows this $\langle \sigma_s\rangle$ vs. $\epsilon$ in a log-log plot. For both values of $V_0$ the difference between the behavior of the hard and soft disks is glaring. In the latter case the importance of the many-body interactions is not decreasing upon approaching jamming, becoming quite independent of $\epsilon$. In hindsight this is not surprising: even when almost touching, two soft colliding spheres $i$ and $j$ which are in the range of interaction of other soft spheres $k, \ell$ etc. should feel their influence; the momentum transferred by the $i,j$ interaction is not determined only by their bare forces, but also by the pull and push of adjacent other disks. In colloquial terms `when push comes to shove it is important who are your neighbors'.

It is interesting to notice that in absolute values the degree of non-mean-field many-body contributions increases when the range of interaction increases (the spheres get softer). The above stated intuition is the obvious
reason for that. But also one should note that the constant value of many-body contributions for the
softer spheres is of the same order as the maximal value of the same contribution for the hard spheres.
There is really very little effect on the proximity of the jamming density on the relative importance
of higher-order forces in the case of softer potentials. One cannot expect that a truncation of the
many-body forces would provide an accurate theory for the critical behavior near jamming.

In summary, it was demonstrated that the hard sphere limit is fragile to softening
in the sense that non-mean-field interactions remain important also in the proximity of the jamming
density. The conclusion is that it is not likely that mean-field calculations in infinite dimensions
would provide accurate predictions for the critical characteristics of jamming in finite dimensions for
generic bare potentials. This conclusion of course does not detract from the relevance of mean field
analytic calculation in indicating the qualitative features of interesting statistical-mechanical
problems, including jamming. Further analysis of the emergent many body interactions in generic
cases and their role in the statistical mechanics of thermal glasses will be discussed in future work.

\acknowledgments
We acknowledge support by the Minerva foundation with
funding from the Federal German Ministry for Education
and Research, the Israel Science Foundation
(Israel Singapore Program) and the Italy-Israel joint laboratory funded by the Ministry
of Foreign affairs. IP is grateful to the Simons Fellowship under the auspices of the
Niels Bohr International Academy.

\clearpage
\begin{center}
	\textbf{\large Supplemental Material to ``How Generic are the Robust Theoretical Aspects of Jamming in Hard Sphere Models?"}
\end{center}

%

\section{Introduction}
In this document we offer Supplementary Information to the main text. In Sect.~\ref{HD} we describe in detail
the creation of jammed configurations of hard disks and their inflation to a wanted density away from jamming.
The next Sect.~\ref{SD} provides similar details for the jamming of the soft disks.
In Sect.~\ref{meta-basin} we describe the manual procedure employed to guarantee that
the mean positions of the particles do not change during our runs. Sect.~\ref{clean} explains the cleaning of the data from the consequences of infrequent collisions. Sect.~\ref{scalar} examines whether computing the inter-particles separation using
a scalar definition may change the main conclusions (it does not). Finally in Sect.~\ref{nonMF} we show that selecting particles with very close-by cage fluctuations does not lead to an elimination of the importance of contributions to the effective forces that
cannot be approximated by the mean-field form.

\section{ Hard Disks simulations }
\label{HD}

To create jammed configurations of hard disks we follow the following steps:
\begin{enumerate}
	
	\item Create systems of harmonic disks at a packing fraction  $\phi = 0.86$.
	
	\item Implement the FIRE minimization algorithm \cite{FIRE} coupled to a Berendsen barostat to bring the pressure to $10^{-6} - 10^{-7}$, depending on the system size.
	
	\item Use 128-bit numerics from here: impose increments of expansive strain that are proportional to the current pressure, and follow each of these increments by a minimization using the FIRE algorithm (without the barostat, i.e. at constant volume). Repeat until the pressure approaches $10^{-10}$ or below.
	
	\item Find the maximal overlap $-h_{ij}$ between any two particles, and expand L precisely to eliminate
	this maximal overlap.
\end{enumerate}
Having determined the system volume as close to the jammed state as possible, we expand the system size from the jammed state by the desired $\epsilon$, cf. Eq.~\ref{defroho} in the main text. (Note that the value of $L_{\rm in}$ fluctuated from realization to realization in our finite size samples). Then the simulation is first equilibrated for $2\times 10^6$ collisions and then run for $n_c=10^8$ collisions more. As described in section \ref{meta-basin}, for each simulation only a section of $n_c/10$ collisions is used for computing the average positions. This is in order to avoid transitions between meta-basins.

For each value of $\epsilon$ the deviation of the force law from the 2-body putative interaction  was averaged over 20-32 configurations, each taken from a different simulation starting from a different initial condition.
\section{Harmonic Disks simulations}
\label{SD}

We use velocity-Verlet  algorithm with time steps $\Delta t=0.001,0.0002$ for $V_0=1000,500000$ respectively to integrate the equation of motion. The Nos\'e-Hoover chain thermostat was used to maintain the desired temperature.

\subsection{Setup of Jammed Configurations:}
To create the jammed configurations, we follow the protocol as described in Refs.~\cite{03OSLN,11SOS}. Starting with a random
configuration in a square box at a temperature $T=0.01$ and low packing fraction of $\phi=0.65$ the system is allowed to equilibrate. Subsequently the system is quenched to a low temperature $T=10^{-12}$, at a rate of $\dot{T} = 10^{-4}$. Finally the energy is minimized (using the conjugate gradient technique).

After reaching the local minimum at initial low packing fraction $\phi_i$, we apply the ``packing finder'' algorithm \cite{03OSLN,11SOS}
to obtain the nearest static packing with infinitesimal particle overlaps. The system is compressed or decompressed, followed by conjugate-gradient energy minimization at each step. Compression is chosen when the total energy is zero
after minimization while decompression is performed when the total energy is nonzero even after energy minimization, due to overlapping
particles. This procedure is terminated when the total potential energy per particle satisfies $U/N < 10^{-16}$ at which point we consider the configuration as jammed. Note that the two methods described
for hard and soft disks appear different but for all practical purposes are in fact equivalent and
could be interchanged.

\subsection{Procedure}

As described in section \ref{meta-basin}, for each simulation only a section of $\Delta \tau=\tau_2/10$ is used for computing the average positions. The aim is to avoid transitions between meta-basins. For each value of $\epsilon$ data was taken from 77-92 configurations each taken from a different simulation starting from a different initial condition.

\section{ Testing changes in meta-basins}
\label{meta-basin}
\begin{figure}
	\includegraphics[scale = 0.50]{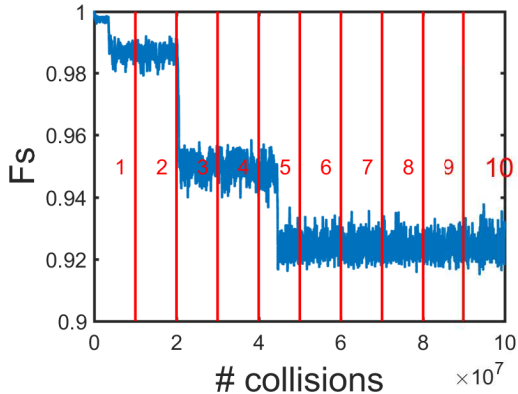}
	\caption{Self-intermediate scattering function measured for a simulation of hard disks at $\epsilon=10^{-3}$. Two transitions between meta-basins are clearly evident as sharp drops. The simulation is divided into 10 temporal sections (delimited by red lines and numbered in the figure) in order to choose a section where no such transitions occur.}
	\label{Fs}
\end{figure}

In order to get reliable average positions, we must guarantee that there are no transitions to different meta-basins during measurements. This is achieved by considering three different time-correlation functions: (i) the self-intermediate scattering function , (ii) the mean square displacement, and (iii) the maximal distance traveled by any single particle.
The time-correlations are measured every 1000 collisions in the hard disks simulations and at each time-step in the harmonic disks simulation.
For hard disks, the values of these correlations functions mostly fluctuate around a constant value  (apart from some initial decay/growth) with infrequent sharp drops/jumps (see Fig. \ref{Fs}). Such a sharp drop/jump in one of these correlation functions indicates a transition between meta-basins. We divide the simulation into 10 temporal sections (with equal number of collisions/ time-steps), and for each simulation analyze a single section in which such transitions were {\em not} observed. All the average positions are computed within such transition-free sections. This is a stricter criterion than the one used in Ref. \cite{16GLPPRR}.
Such sharp drops are  observed mostly for the larger  expansions $\epsilon\ge10^{-4}$. For smaller expansions the simulation time is too short for transitions to occur.
For the soft harmonic disks, the values of the correlations functions can also change smoothly and  we choose simulation sections where these values fluctuate around constant values without observable decay or growth.

\section { Cleaning the data from infrequent collisions}
\label{clean}
\begin{figure}
	\includegraphics[scale = 0.40]{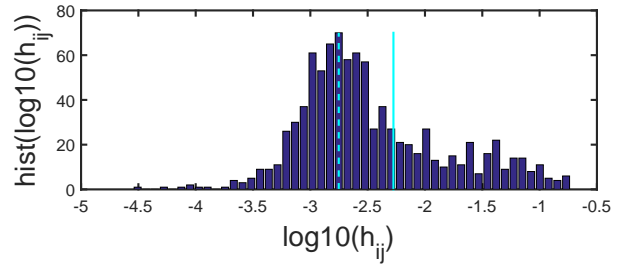}
	\caption{Histogram of $log_{10}(h_{ij})$ for a hard disks configuration at expansion $\epsilon=10^{-3}$. Dashed cyan line indicates the most frequent bin $h_{ij}^{freq}$. The analysis employs effective forces $\B f_{ij}$ for which $h_{ij} \le 3 \times h_{ij}^{freq}$ (solid cyan line).}
	\label{cleanfig}
\end{figure}

{\bf Hard Disks:}
Configurations of hard disks involve ``rattlers" that collide only infrequently compared
to typical disks. This intoduces errors in the effective force measurements.
To clean the data from such outliers we identify the range of $h_{ij}$ that can be trusted.
To this aim  we plot a histogram of $log_{10}(h_{ij})$ and bin it into 50 bins, cf. Fig.~\ref{cleanfig}. The value of $h_{ij}$ in the bin with the
highest weight is denoted as $h_{ij}^{freq}$. We then include only effective forces $\B f_{ij}$ for which $h_{ij}\le 3 \times h_{ij}^{freq}$.

{\bf Harmonic Disks:}
In the case of harmonic disks the ``gaps" $h_{ij}$ can be negative, and the cleaning
of the data is a bit more tricky. Instead of using the gap  $ h_{ij}=r_{ij}-\sigma_i-\sigma_j$, we used $\tilde h_{ij}=r_{ij}-r_{ij}^{min}$ where $r_{ij}^{min}$, is the minimal $r_{ij}$ of the relevant interaction. We then followed the same procedure as for the hard disks:  We plotted a histogram (50 bins) of $log_{10}(\tilde h_{ij})$, found the largest bin $\tilde h_{ij}^{freq}$ and considered effective forces associated with $\tilde h_{ij}\le 3 \times \tilde h_{ij}^{freq}$.

\begin{figure}
	\includegraphics[scale = 0.60]{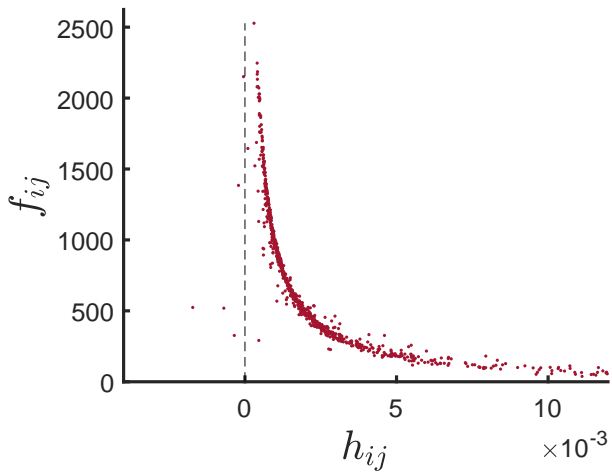}
	\caption{The effective forces in the hard sphere case with $\epsilon=10^{-3}$. The
		data is shown only for small $h_{ij}$ to provide higher resolution around the rare events with negative $h_{ij}$. The few negative values of $h_{ij}$ are real, stemming from dynamics in which the difference in {\em average} positions are indeed negative. }
	\label{fulldata}
\end{figure}

Besides cleaning the data from pairs having very large values of $h_{ij}$, one should also consider for both soft and hard disks some rare configurations that include particle pairs with extremely small and negative values of $h_{ij}$ that deviate strongly from the typical behavior, exhibiting abnormally small forces $f_{ij}$. These abnormally small $f_{ij}$ were not considered in the analysis. This rare phenomenon disappears when the definition of $h_{ij}$
is changed in favor of a scalar average, and see Sect.~\ref{scalar} below. At any rate these
rare events do not change the general conclusions of the study, as is shown explicitly in Sect.~\ref{scalar}. To get an impression of the data before the clean-up of negative $h_{ij}$ we present in
Fig.~\ref{fulldata} some of the effective forces computed for the hard disk case as a function of $h_{ij}$. It is visually clear that the problematic points are rare.

\begin{figure}
	\includegraphics[scale = 0.50]{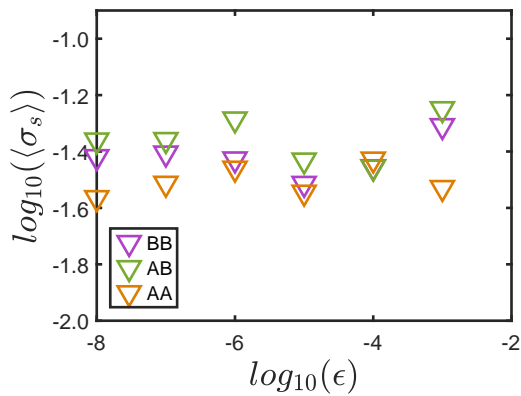}
	\caption{The standard deviation from the mean field binary force law for soft spheres as a function of the distance from jamming. Here $V_0=1000$. Here we used the scalar definition of the distance between paires and particles and we show the results for each
		type of interaction (AA, BB and AB) separately for extra care.}
	\label{examp}
\end{figure}
\begin{figure}
	\includegraphics[scale = 0.50]{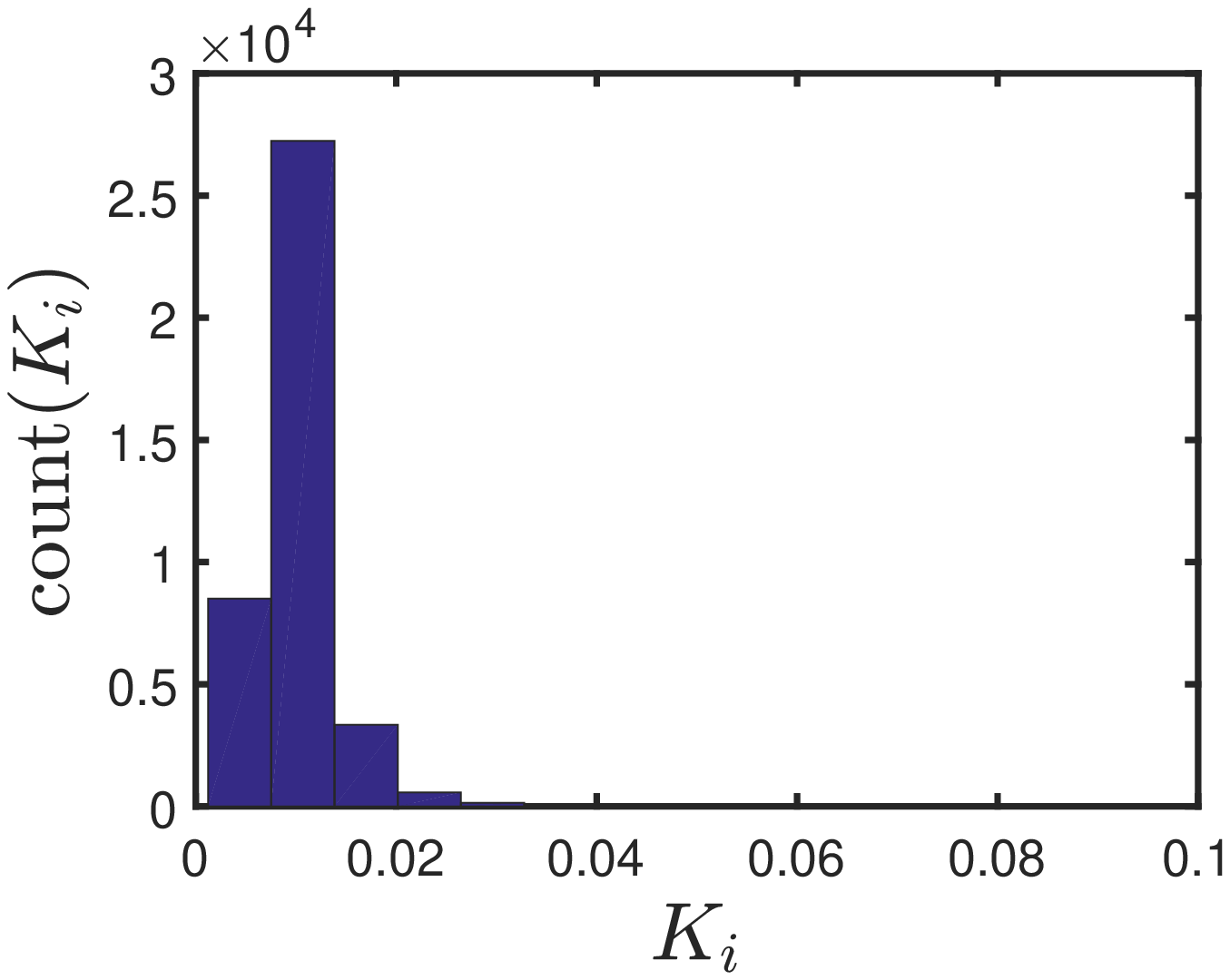}
	\includegraphics[scale = 0.50]{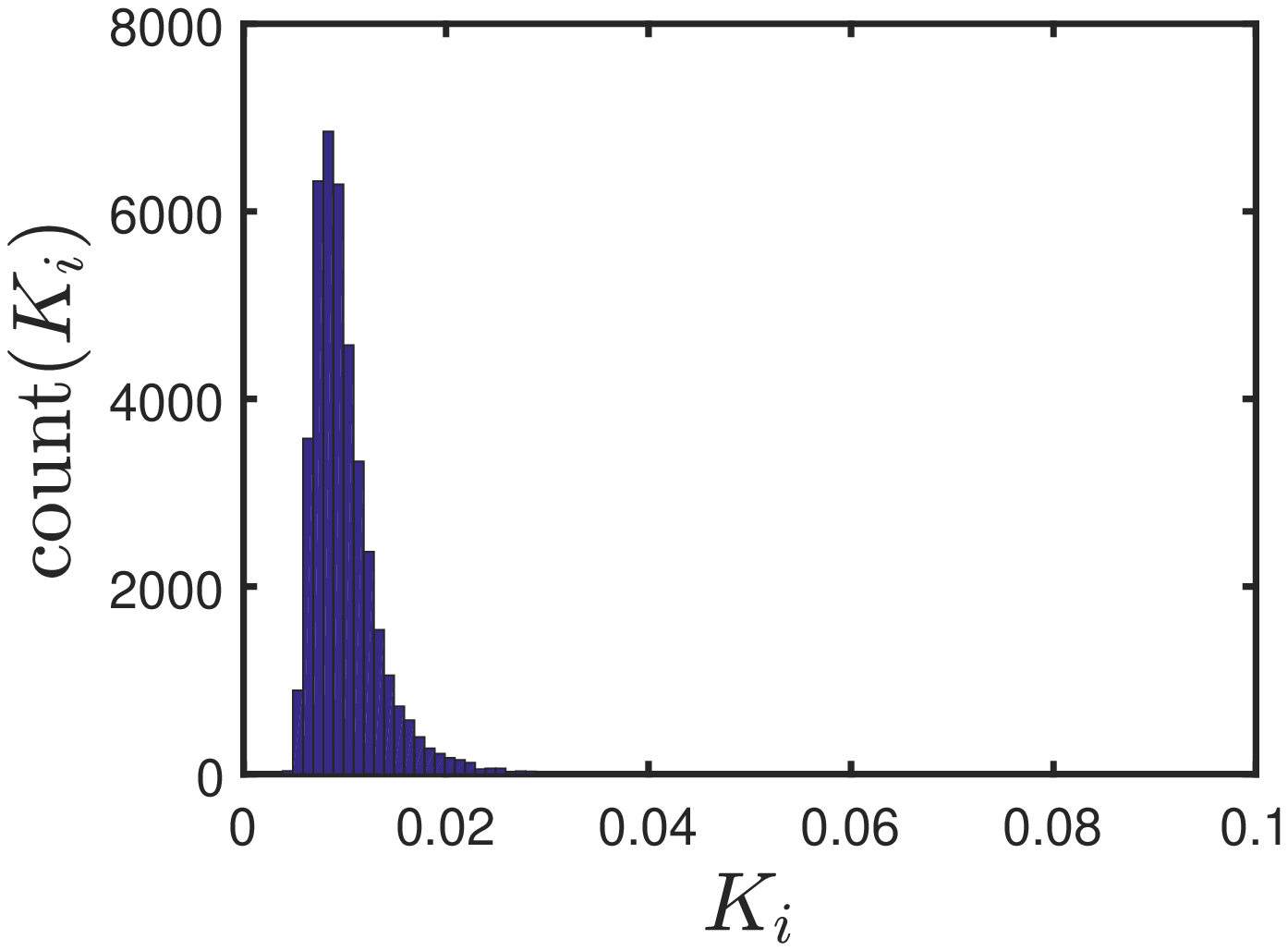}
	\includegraphics[scale = 0.50]{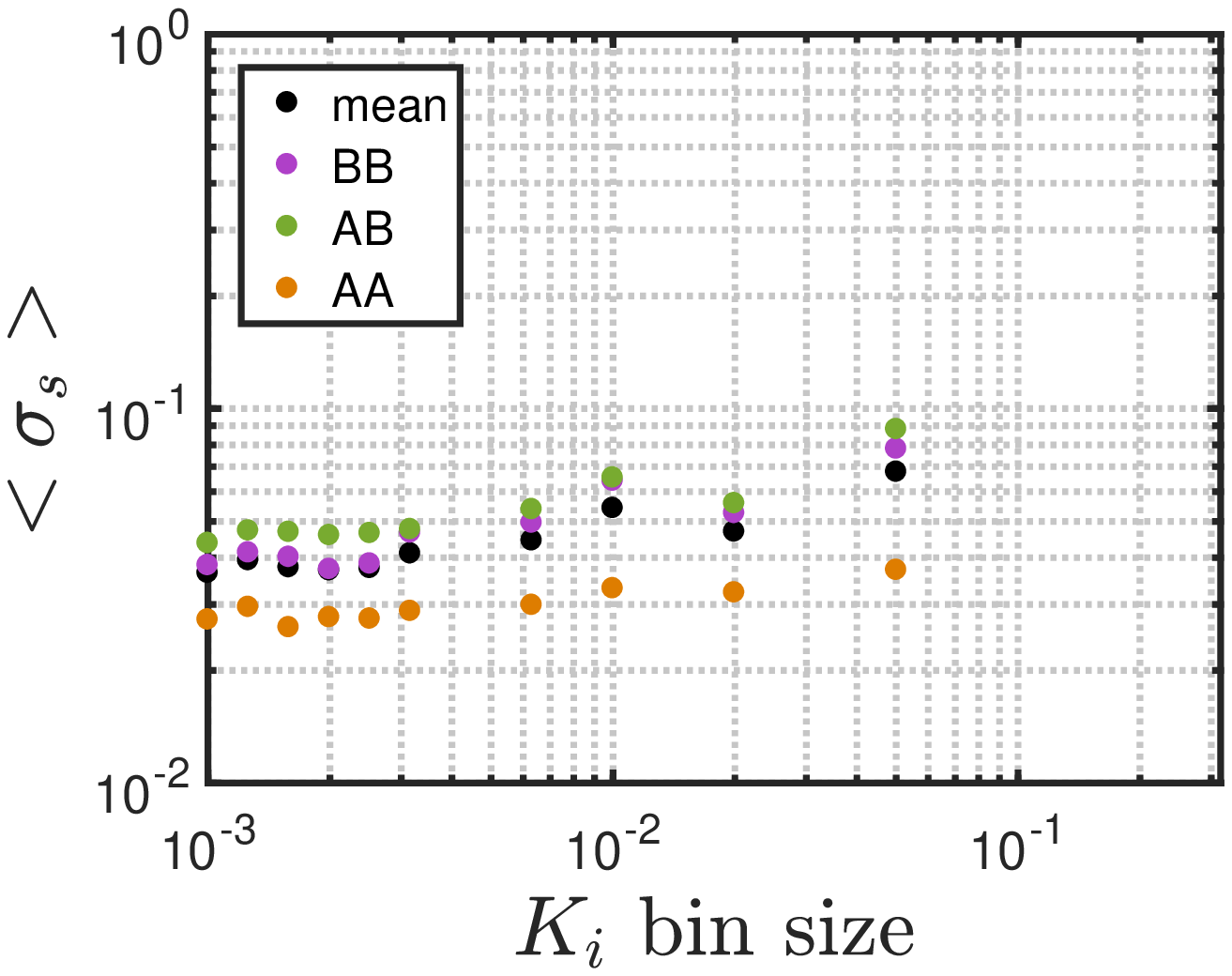}
	\caption{Upper panel: the histogram of values of $K_i$ with coarse bins. Middle panel: the same histogram with finer bins. Lower panel: The standard deviation around a binary force law as a function of the bin size.}
	\label{nomean}
\end{figure}

\section{analysis with scalar averaging of distances $r_{ij}$}
\label{scalar}
Instead of using the definition of $h_{ij}$ in which $\bar r_{ij}\equiv |\frac{1}{\tau}\int_0^\tau
dt ~\B r_{ij}(t) |$ which is computed as a vector average,
one could employ a scalar definition of the mean distance
between particles,
\begin{equation}
\tilde h_{ij} =  \tilde r_{ij} -R_i - R_j  \ ; \quad \tilde r_{ij} \equiv \frac{1}{\tau}\int_0^\tau
dt~ r_{ij}(t) \ .
\end{equation}
For the case of soft spheres we checked carefully whether this definition may lead
to a different conclusion. The answer is negative. As an example we show in Fig.~\ref{examp}
the computed contribution of many body interactions as a function of $\epsilon$. The overall
order of magnitude of the standard deviation reduces compared to the vector definition of the
distances, but still there is no indication for
approaching the binary limit when $\epsilon\to 0$.

\section{Ruling out mean field effective forces in soft spheres}
\label{nonMF}
In order to determine whether in a given system the force-law conforms with mean-field expectations we need to determine the cage fluctuations $K_i$.
The probability distribution function (pdf) of $K_i$ was measured for soft spheres using some 77-92 configurations (depending on $\epsilon$). Next, we selected pairs of particles from a bin
of $K_i$ value with decreasing width of the bin. If Eq.~\ref{MF} of the main text pertains, we should
expect that reducing the bin size and plotting the effective forces as a function of $h_{ij}$ must
result in reducing the scatter around a functional behavior. In Fig.~\ref{nomean} we show that
this is not the case. The data shown pertains to particle pairs whose $K_i\approx K_j$ up to the
bin width, selected from the bin with highest weight. In the upper panel we show
the histogram of $K_i$ with large bins, and in the middle panel with finer bins. Finally, in the lower panel we show that the contribution of non-binary interaction does not reduce when the bins of the histogram get finer and finer. The conclusion is that the mean-field expectation Eq.~\ref{MF} of the main text is untenable in the case of
harmonic spheres.

\bibliography{LJ}

\begin{thebibliography}{21}%
\makeatletter
\providecommand \@ifxundefined [1]{%
 \@ifx{#1\undefined}
}%
\providecommand \@ifnum [1]{%
 \ifnum #1\expandafter \@firstoftwo
 \else \expandafter \@secondoftwo
 \fi
}%
\providecommand \@ifx [1]{%
 \ifx #1\expandafter \@firstoftwo
 \else \expandafter \@secondoftwo
 \fi
}%
\providecommand \natexlab [1]{#1}%
\providecommand \enquote  [1]{``#1''}%
\providecommand \bibnamefont  [1]{#1}%
\providecommand \bibfnamefont [1]{#1}%
\providecommand \citenamefont [1]{#1}%
\providecommand \href@noop [0]{\@secondoftwo}%
\providecommand \href [0]{\begingroup \@sanitize@url \@href}%
\providecommand \@href[1]{\@@startlink{#1}\@@href}%
\providecommand \@@href[1]{\endgroup#1\@@endlink}%
\providecommand \@sanitize@url [0]{\catcode `\\12\catcode `\$12\catcode
  `\&12\catcode `\#12\catcode `\^12\catcode `\_12\catcode `\%12\relax}%
\providecommand \@@startlink[1]{}%
\providecommand \@@endlink[0]{}%
\providecommand \url  [0]{\begingroup\@sanitize@url \@url }%
\providecommand \@url [1]{\endgroup\@href {#1}{\urlprefix }}%
\providecommand \urlprefix  [0]{URL }%
\providecommand \Eprint [0]{\href }%
\providecommand \doibase [0]{http://dx.doi.org/}%
\providecommand \selectlanguage [0]{\@gobble}%
\providecommand \bibinfo  [0]{\@secondoftwo}%
\providecommand \bibfield  [0]{\@secondoftwo}%
\providecommand \translation [1]{[#1]}%
\providecommand \BibitemOpen [0]{}%
\providecommand \bibitemStop [0]{}%
\providecommand \bibitemNoStop [0]{.\EOS\space}%
\providecommand \EOS [0]{\spacefactor3000\relax}%
\providecommand \BibitemShut  [1]{\csname bibitem#1\endcsname}%
\let\auto@bib@innerbib\@empty
\bibitem [{\citenamefont {Hansen}\ and\ \citenamefont {McDonald}(2006)}]{06HM}%
  \BibitemOpen
  \bibfield  {author} {\bibinfo {author} {\bibfnamefont {J.}~\bibnamefont
  {Hansen}}\ and\ \bibinfo {author} {\bibfnamefont {I.}~\bibnamefont
  {McDonald}},\ }in\ \href {\doibase
  http://doi.org/10.1016/B978-012370535-8/50000-8} {\emph {\bibinfo {booktitle}
  {Theory of Simple Liquids (Third Edition)}}},\ \bibinfo {editor} {edited by\
  \bibinfo {editor} {\bibfnamefont {J.-P.}\ \bibnamefont {Hansen}}, , \ and\
  \bibinfo {editor} {\bibfnamefont {I.~R.}\ \bibnamefont {McDonald}}}\
  (\bibinfo  {publisher} {Academic Press},\ \bibinfo {address} {Burlington},\
  \bibinfo {year} {2006})\BibitemShut {NoStop}%
\bibitem [{\citenamefont {Rapaport}(1997)}]{99Rap}%
  \BibitemOpen
  \bibfield  {author} {\bibinfo {author} {\bibfnamefont {D.}~\bibnamefont
  {Rapaport}},\ }\href {https://books.google.com/books?id=hHxllwEACAAJ} {\emph
  {\bibinfo {title} {The Art of Molecular Dynamics Simulation}}}\ (\bibinfo
  {publisher} {Cambridge University Press},\ \bibinfo {year}
  {1997})\BibitemShut {NoStop}%
\bibitem [{\citenamefont {O'Hern}\ \emph {et~al.}(2003)\citenamefont {O'Hern},
  \citenamefont {Silbert}, \citenamefont {Liu},\ and\ \citenamefont
  {Nagel}}]{03OSLN}%
  \BibitemOpen
  \bibfield  {author} {\bibinfo {author} {\bibfnamefont {C.~S.}\ \bibnamefont
  {O'Hern}}, \bibinfo {author} {\bibfnamefont {L.~E.}\ \bibnamefont {Silbert}},
  \bibinfo {author} {\bibfnamefont {A.~J.}\ \bibnamefont {Liu}}, \ and\
  \bibinfo {author} {\bibfnamefont {S.~R.}\ \bibnamefont {Nagel}},\ }\href
  {\doibase 10.1103/PhysRevE.68.011306} {\bibfield  {journal} {\bibinfo
  {journal} {Phys. Rev. E}\ }\textbf {\bibinfo {volume} {68}},\ \bibinfo
  {pages} {011306} (\bibinfo {year} {2003})}\BibitemShut {NoStop}%
\bibitem [{\citenamefont {Silbert}\ \emph {et~al.}(2002)\citenamefont
  {Silbert}, \citenamefont {Ertas}, \citenamefont {Grest}, \citenamefont
  {Halsey},\ and\ \citenamefont {Levine}}]{02SEGHL}%
  \BibitemOpen
  \bibfield  {author} {\bibinfo {author} {\bibfnamefont {L.~E.}\ \bibnamefont
  {Silbert}}, \bibinfo {author} {\bibfnamefont {D.}~\bibnamefont {Ertas}},
  \bibinfo {author} {\bibfnamefont {G.~S.}\ \bibnamefont {Grest}}, \bibinfo
  {author} {\bibfnamefont {T.~C.}\ \bibnamefont {Halsey}}, \ and\ \bibinfo
  {author} {\bibfnamefont {D.}~\bibnamefont {Levine}},\ }\href {\doibase
  10.1103/PhysRevE.65.031304} {\bibfield  {journal} {\bibinfo  {journal} {Phys.
  Rev. E}\ }\textbf {\bibinfo {volume} {65}},\ \bibinfo {pages} {031304}
  (\bibinfo {year} {2002})}\BibitemShut {NoStop}%
\bibitem [{\citenamefont {O'Hern}\ \emph {et~al.}(2002)\citenamefont {O'Hern},
  \citenamefont {Langer}, \citenamefont {Liu},\ and\ \citenamefont
  {Nagel}}]{02OLLN}%
  \BibitemOpen
  \bibfield  {author} {\bibinfo {author} {\bibfnamefont {C.~S.}\ \bibnamefont
  {O'Hern}}, \bibinfo {author} {\bibfnamefont {S.~A.}\ \bibnamefont {Langer}},
  \bibinfo {author} {\bibfnamefont {A.~J.}\ \bibnamefont {Liu}}, \ and\
  \bibinfo {author} {\bibfnamefont {S.~R.}\ \bibnamefont {Nagel}},\ }\href
  {\doibase 10.1103/PhysRevLett.88.075507} {\bibfield  {journal} {\bibinfo
  {journal} {Phys. Rev. Lett.}\ }\textbf {\bibinfo {volume} {88}},\ \bibinfo
  {pages} {075507} (\bibinfo {year} {2002})}\BibitemShut {NoStop}%
\bibitem [{\citenamefont {Berthier}\ \emph {et~al.}(2016)\citenamefont
  {Berthier}, \citenamefont {Coslovich}, \citenamefont {Ninarello},\ and\
  \citenamefont {Ozawa}}]{16BCNO}%
  \BibitemOpen
  \bibfield  {author} {\bibinfo {author} {\bibfnamefont {L.}~\bibnamefont
  {Berthier}}, \bibinfo {author} {\bibfnamefont {D.}~\bibnamefont {Coslovich}},
  \bibinfo {author} {\bibfnamefont {A.}~\bibnamefont {Ninarello}}, \ and\
  \bibinfo {author} {\bibfnamefont {M.}~\bibnamefont {Ozawa}},\ }\href
  {\doibase 10.1103/PhysRevLett.116.238002} {\bibfield  {journal} {\bibinfo
  {journal} {Phys. Rev. Lett.}\ }\textbf {\bibinfo {volume} {116}},\ \bibinfo
  {pages} {238002} (\bibinfo {year} {2016})}\BibitemShut {NoStop}%
\bibitem [{\citenamefont {Charbonneau}\ \emph {et~al.}(2014)\citenamefont
  {Charbonneau}, \citenamefont {Kurchan}, \citenamefont {Parisi}, \citenamefont
  {Urabni},\ and\ \citenamefont {Zamponi}}]{14CKPUZ}%
  \BibitemOpen
  \bibfield  {author} {\bibinfo {author} {\bibfnamefont {P.}~\bibnamefont
  {Charbonneau}}, \bibinfo {author} {\bibfnamefont {J.}~\bibnamefont
  {Kurchan}}, \bibinfo {author} {\bibfnamefont {G.}~\bibnamefont {Parisi}},
  \bibinfo {author} {\bibfnamefont {P.}~\bibnamefont {Urabni}}, \ and\ \bibinfo
  {author} {\bibfnamefont {F.}~\bibnamefont {Zamponi}},\ }\href@noop {}
  {\bibfield  {journal} {\bibinfo  {journal} {Nature Communications}\ }\textbf
  {\bibinfo {volume} {5}},\ \bibinfo {pages} {3725} (\bibinfo {year}
  {2014})}\BibitemShut {NoStop}%
\bibitem [{\citenamefont {Charbonneau}\ \emph {et~al.}(2015)\citenamefont
  {Charbonneau}, \citenamefont {Corwin}, \citenamefont {Parisi},\ and\
  \citenamefont {Zamponi}}]{15CCPZ}%
  \BibitemOpen
  \bibfield  {author} {\bibinfo {author} {\bibfnamefont {P.}~\bibnamefont
  {Charbonneau}}, \bibinfo {author} {\bibfnamefont {E.}~\bibnamefont {Corwin}},
  \bibinfo {author} {\bibfnamefont {G.}~\bibnamefont {Parisi}}, \ and\ \bibinfo
  {author} {\bibfnamefont {F.}~\bibnamefont {Zamponi}},\ }\href@noop {}
  {\bibfield  {journal} {\bibinfo  {journal} {Phys. Rev. Lett.}\ }\textbf
  {\bibinfo {volume} {114}},\ \bibinfo {pages} {125504} (\bibinfo {year}
  {2015})}\BibitemShut {NoStop}%
\bibitem [{\citenamefont {Berthier}\ \emph {et~al.}(2011)\citenamefont
  {Berthier}, \citenamefont {Jacquin},\ and\ \citenamefont {Zamponi}}]{11BJZ}%
  \BibitemOpen
  \bibfield  {author} {\bibinfo {author} {\bibfnamefont {L.}~\bibnamefont
  {Berthier}}, \bibinfo {author} {\bibfnamefont {H.}~\bibnamefont {Jacquin}}, \
  and\ \bibinfo {author} {\bibfnamefont {F.}~\bibnamefont {Zamponi}},\
  }\href@noop {} {\bibfield  {journal} {\bibinfo  {journal} {Phys. Rev. E}\ }
  (\bibinfo {year} {2011})}\BibitemShut {NoStop}%
\bibitem [{\citenamefont {Charbonneau}\ \emph {et~al.}(2017)\citenamefont
  {Charbonneau}, \citenamefont {Kurchan}, \citenamefont {Parisi}, \citenamefont
  {Urbani},\ and\ \citenamefont {Zamponi}}]{17CKPUZ}%
  \BibitemOpen
  \bibfield  {author} {\bibinfo {author} {\bibfnamefont {P.}~\bibnamefont
  {Charbonneau}}, \bibinfo {author} {\bibfnamefont {J.}~\bibnamefont
  {Kurchan}}, \bibinfo {author} {\bibfnamefont {G.}~\bibnamefont {Parisi}},
  \bibinfo {author} {\bibfnamefont {P.}~\bibnamefont {Urbani}}, \ and\ \bibinfo
  {author} {\bibfnamefont {F.}~\bibnamefont {Zamponi}},\ }\href@noop {}
  {\bibfield  {journal} {\bibinfo  {journal} {Annual Review of Condensed Matter
  Physics}\ }\textbf {\bibinfo {volume} {8}},\ \bibinfo {pages} {265} (\bibinfo
  {year} {2017})}\BibitemShut {NoStop}%
\bibitem [{\citenamefont {Parisi}\ and\ \citenamefont {Zamponi}(2006)}]{06PZ}%
  \BibitemOpen
  \bibfield  {author} {\bibinfo {author} {\bibfnamefont {G.}~\bibnamefont
  {Parisi}}\ and\ \bibinfo {author} {\bibfnamefont {F.}~\bibnamefont
  {Zamponi}},\ }\href {http://stacks.iop.org/1742-5468/2006/i=03/a=P03017}
  {\bibfield  {journal} {\bibinfo  {journal} {Journal of Statistical Mechanics:
  Theory and Experiment}\ }\textbf {\bibinfo {volume} {2006}},\ \bibinfo
  {pages} {P03017} (\bibinfo {year} {2006})}\BibitemShut {NoStop}%
\bibitem [{\citenamefont {Brito}\ and\ \citenamefont {Wyart}(2006)}]{06BW}%
  \BibitemOpen
  \bibfield  {author} {\bibinfo {author} {\bibfnamefont {C.}~\bibnamefont
  {Brito}}\ and\ \bibinfo {author} {\bibfnamefont {M.}~\bibnamefont {Wyart}},\
  }\href {http://stacks.iop.org/0295-5075/76/i=1/a=149} {\bibfield  {journal}
  {\bibinfo  {journal} {EPL (Europhysics Letters)}\ }\textbf {\bibinfo {volume}
  {76}},\ \bibinfo {pages} {149} (\bibinfo {year} {2006})}\BibitemShut
  {NoStop}%
\bibitem [{\citenamefont {Gendelman}\ \emph
  {et~al.}(2016{\natexlab{a}})\citenamefont {Gendelman}, \citenamefont
  {Lerner}, \citenamefont {Pollack}, \citenamefont {Procaccia}, \citenamefont
  {Rainone},\ and\ \citenamefont {Riechers}}]{16GLPPRR}%
  \BibitemOpen
  \bibfield  {author} {\bibinfo {author} {\bibfnamefont {O.}~\bibnamefont
  {Gendelman}}, \bibinfo {author} {\bibfnamefont {E.}~\bibnamefont {Lerner}},
  \bibinfo {author} {\bibfnamefont {Y.~G.}\ \bibnamefont {Pollack}}, \bibinfo
  {author} {\bibfnamefont {I.}~\bibnamefont {Procaccia}}, \bibinfo {author}
  {\bibfnamefont {C.}~\bibnamefont {Rainone}}, \ and\ \bibinfo {author}
  {\bibfnamefont {B.}~\bibnamefont {Riechers}},\ }\href {\doibase
  10.1103/PhysRevE.94.051001} {\bibfield  {journal} {\bibinfo  {journal} {Phys.
  Rev. E}\ }\textbf {\bibinfo {volume} {94}},\ \bibinfo {pages} {051001}
  (\bibinfo {year} {2016}{\natexlab{a}})}\BibitemShut {NoStop}%
\bibitem [{\citenamefont {Gendelman}\ \emph
  {et~al.}(2016{\natexlab{b}})\citenamefont {Gendelman}, \citenamefont
  {Pollack},\ and\ \citenamefont {Procaccia}}]{16GPP}%
  \BibitemOpen
  \bibfield  {author} {\bibinfo {author} {\bibfnamefont {O.}~\bibnamefont
  {Gendelman}}, \bibinfo {author} {\bibfnamefont {Y.~G.}\ \bibnamefont
  {Pollack}}, \ and\ \bibinfo {author} {\bibfnamefont {I.}~\bibnamefont
  {Procaccia}},\ }\href {\doibase 10.1103/PhysRevE.93.060601} {\bibfield
  {journal} {\bibinfo  {journal} {Phys. Rev. E}\ }\textbf {\bibinfo {volume}
  {93}},\ \bibinfo {pages} {060601} (\bibinfo {year}
  {2016}{\natexlab{b}})}\BibitemShut {NoStop}%
\bibitem [{\citenamefont {Plefka}(1982)}]{82P}%
  \BibitemOpen
  \bibfield  {author} {\bibinfo {author} {\bibfnamefont {T.}~\bibnamefont
  {Plefka}},\ }\href@noop {} {\bibfield  {journal} {\bibinfo  {journal}
  {Journal of Physics A: Mathematical and general}\ }\textbf {\bibinfo {volume}
  {15}},\ \bibinfo {pages} {1971} (\bibinfo {year} {1982})}\BibitemShut
  {NoStop}%
\bibitem [{\citenamefont {Georges}\ and\ \citenamefont {Yedidia}(1991)}]{91GY}%
  \BibitemOpen
  \bibfield  {author} {\bibinfo {author} {\bibfnamefont {A.}~\bibnamefont
  {Georges}}\ and\ \bibinfo {author} {\bibfnamefont {J.}~\bibnamefont
  {Yedidia}},\ }\href@noop {} {\bibfield  {journal} {\bibinfo  {journal}
  {Journal of Physics A: Mathematical and General}\ }\textbf {\bibinfo {volume}
  {24}},\ \bibinfo {pages} {2173} (\bibinfo {year} {1991})}\BibitemShut
  {NoStop}%
\bibitem [{Note1()}]{Note1}%
  \BibitemOpen
  \bibinfo {note} {In principle one can also consider a dependence on a
  tensorial quantity $K^{\alpha \beta }_i\equiv \protect \sqrt {\delimiter
  "426830A ( r^\alpha _i(t)-\protect \mathaccentV {bar}016{r}^\alpha _i)(
  r^\beta _i(t)-\protect \mathaccentV {bar}016{r}^\beta _i)\delimiter "526930B
  }$}\BibitemShut {NoStop}%
\bibitem [{\citenamefont {Schreck}\ \emph {et~al.}(2011)\citenamefont
  {Schreck}, \citenamefont {O'Hern},\ and\ \citenamefont {Silbert}}]{11SOS}%
  \BibitemOpen
  \bibfield  {author} {\bibinfo {author} {\bibfnamefont {C.~F.}\ \bibnamefont
  {Schreck}}, \bibinfo {author} {\bibfnamefont {C.~S.}\ \bibnamefont {O'Hern}},
  \ and\ \bibinfo {author} {\bibfnamefont {L.~E.}\ \bibnamefont {Silbert}},\
  }\href {\doibase 10.1103/PhysRevE.84.011305} {\bibfield  {journal} {\bibinfo
  {journal} {Phys. Rev. E}\ }\textbf {\bibinfo {volume} {84}},\ \bibinfo
  {pages} {011305} (\bibinfo {year} {2011})}\BibitemShut {NoStop}%
\bibitem [{Note2()}]{Note2}%
  \BibitemOpen
  \bibinfo {note} {See in the Supplemental Material another, scalar definition
  of the average gap and its consequences}\BibitemShut {NoStop}%
\bibitem [{Note3()}]{Note3}%
  \BibitemOpen
  \bibinfo {note} {Note that in the case of soft spheres we computed the
  standard deviation from all the samples together, whereas in the hard spheres
  we averaged over individual configurations. Doing the same in the present
  introduces no difference.}\BibitemShut {Stop}%
\bibitem [{\citenamefont {Bitzek}\ \emph {et~al.}(2006)\citenamefont {Bitzek},
  \citenamefont {Koskinen}, \citenamefont {G\"ahler}, \citenamefont {Moseler},\
  and\ \citenamefont {Gumbsch}}]{FIRE}%
  \BibitemOpen
  \bibfield  {author} {\bibinfo {author} {\bibfnamefont {E.}~\bibnamefont
  {Bitzek}}, \bibinfo {author} {\bibfnamefont {P.}~\bibnamefont {Koskinen}},
  \bibinfo {author} {\bibfnamefont {F.}~\bibnamefont {G\"ahler}}, \bibinfo
  {author} {\bibfnamefont {M.}~\bibnamefont {Moseler}}, \ and\ \bibinfo
  {author} {\bibfnamefont {P.}~\bibnamefont {Gumbsch}},\ }\href {\doibase
  10.1103/PhysRevLett.97.170201} {\bibfield  {journal} {\bibinfo  {journal}
  {Phys. Rev. Lett.}\ }\textbf {\bibinfo {volume} {97}},\ \bibinfo {pages}
  {170201} (\bibinfo {year} {2006})}\BibitemShut {NoStop}%
\end{thebibliography}%
\end{document}